\documentclass[a4paper,10pt]{article}
\usepackage[dvips]{graphicx}
\usepackage{amssymb,amsmath}
\numberwithin{equation}{section}

\begin{document}

\title{FRW  cosmological models   with integrable and nonintegrable  differential equations of state}
\vspace{4cm}
\author{ Shynaray Myrzakul$^1$,  Kuralay Esmakhanova$^1$, Kairat Myrzakulov$^1$,  \\Gulgasyl  Nugmanova$^1$ and        Ratbay Myrzakulov$^{1,2,}$\footnote{The corresponding author. Email: rmyrzakulov@gmail.com; rmyrzakulov@csufresno.edu}\vspace{1cm}\\ \textit{$^1$Eurasian International Center for Theoretical Physics,} \\ \textit{Eurasian National University, Astana 010008, Kazakhstan}\\ \textit{$^2$Department of Physics, CSU Fresno, Fresno, CA 93740 USA}}

\date{}

\maketitle
\begin{abstract} 
The basic models of modern cosmology work in  FRW spacetime. Hence follows that it is important  to study the physical and mathematical nature of FRW cosmological models. In this work,  we consider FRW models with the differential equations of state. Some  new classes integrable and nonintegrable FRW cosmological models   were constructed.  It is remarkable that all proposed integrable and nonintegrable FRW models admit exact solutions. For some of them, such exact solutions are presented. Some artificial two-dimensional FRW models were  also proposed. Finally in Appendix, we extend  the obtained results for g-essence models and for its two reductions: k-essence and f-essence. 
\end{abstract}
\sloppy
\tableofcontents
\section{Introduction} 
 FRW models play a central role in modern cosmology. In particular, almost all popular theoretical models of dark energy work in FRW spacetime. Again almost all models of dark energy meet some difficulties like cosmological constant problems,  fine-tuning problems and so on. One of consequences of such difficulties of modern cosmology  is the necessity  more carefully investigate the basics of General Relativity (GR), in  particular, FRW cosmology. One of the poorly studied sector of modern theoretical cosmology is the problem of  integrability of cosmological  models, especially, FRW cosmological models. In our previous papers \cite{Kuralay1}-\cite{Gulia}, we have studied the relationship between the basic cosmological equations, namely, the Friedmann equations
 \begin{eqnarray}
		p&=&-2\dot{H}-3H^2,\\
	\rho&=&3H^2
	\end{eqnarray}
	and some important equations known from the other branches of physics and/or mathematics. These last equations include:
	\\
	a) The Ramanujan equation
	\begin{eqnarray}
	\dot{F}&=&\frac{F^2-E}{12},\\ 
		\dot{E}&=&\frac{FE-J}{3},\\
			\dot{J}&=&\frac{FJ- E^2}{2}.
	\end{eqnarray} 
	\\
	b) The Chazy-III equation
		\begin{equation}
\dddot{y}=2y\ddot{y}-3\dot{y}^{2}.
\end{equation}
	c) The Lorenz  oscillator equation
	\begin{eqnarray}
X_N&=&\sigma(Y-X),\\ 
	Y_N&=&X(\delta-Z)-Y,\\
Z_N&=&XY-\beta Z.
	\end{eqnarray}
	d) Painlev$\acute{e}$ equations. 	These six equations, traditionally called Painlev$\acute{e}$ I-VI, are as follows:
\begin{equation}
 \begin{array}{|c|c|}
\hline \\
P_{I} - equation &   \ddot{y}=6y^2+t  \\
\hline \\
P_{II} - equation &  \ddot{y}=2y^3+t y+\alpha  \\
\hline \\
P_{III} - equation &  \ddot{y}=\frac{1}{y}\dot{y}^2-\frac{1}{t}(\dot{y}-\alpha y^2-\beta)+\gamma y^3+\frac{\delta}{y} \\
\hline \\
P_{IV} - equation &  \ddot{y}=\frac{1}{2y}\dot{y}^2+1.5y^3+4t y^2+2(t^2-\alpha)y+
 \frac{\delta}{y} \\
\hline \\
P_{V} - equation &  \ddot{y}=(\frac{1}{2y}+\frac{1}{y-1})\dot{y}^2
 -\frac{1}{t}(\dot{y}-\gamma  y)+\frac{(y-1)^2}{t^{2}}(\alpha y+\frac{\beta}{y})
 +\frac{\delta y(y+1)}{y-1} \\
\hline \\
P_{VI} - equation &  \ddot{y}=\varphi(t)\dot{y}^2
 -\xi(t)\dot{y}
 +\frac{y(y-1)(y-t)}{t^{2}(t-1)^{2}}[\alpha+ \frac{\beta t}{ y^{2}}+\zeta(t)]\\
\hline
\end{array} \end{equation}
Here $\varphi(z)=0.5[y^{-1}+(y-1)^{-1}+(y-z)^{-1}], \quad \xi(z)= [z^{-1}+(z-1)^{-1}+(y-z)^{-1}],\quad \zeta(z)=[\gamma(z-1)(y-1)^{-2}+\delta z(z-1)(y-z)^{-2}]$.

  In this work, we study the FRW cosmological models with the differential equations of state (EoS). To construct   integrable and nonintegrable reductions of such FRW cosmological  models we use  the implantation method.  In particular, to construct integrable FRW models in one dimensions we use six Painlev$\acute{e}$ equations (see also \cite{Kuralay1}-\cite{Gulia}).

The paper is organized as follows. In section 2, we  present the basic equations of FRW cosmology. Section 3 is devoted to study the FRW models with the half EoF and Section 4 to the FRW models with the full EoS.  In the next section 5, we present the so-called D - models. Some artificial two-dimensional models were constructed in the section 6. The last section 7 is  devoted to the conclusion.
\section{Basic gravitational  equations}
We start  from the classical GR case.  In this GR case, the standard gravitational action has the form
  \begin{equation}
S=\int\sqrt{-g}d^4x (R+L_m-\Lambda), \end{equation}
where $R$ is the scalar curvature and $L_m$ is the Lagrangian of the matter. 
We work with the FRW spacetime which has  the  metric
   \begin{equation}
ds^2=-dt^2+a^2(t)\left[\frac{dr^2}{1-kr^2}+r^2d\Omega^2\right], \end{equation}
where $a(t)$ is the scale factor, $k=-1,0,1$ represent the three-dimensional space with the negative, zero, and positive spatial curvature, respectively.   In this case the  Ricci scalar reads as
   \begin{equation}
  R=6\left[\frac{\ddot{a}}{a}+\left(\frac{\dot{a}}{a}\right)^2+\frac{k}{a^2}\right],  \end{equation}
 where a dot represents differentiation with respect to $t$. The Friedmann equations read as
  \begin{equation}
  \left(\frac{\dot{a}}{a}\right)^2=\frac{\Lambda}{3}-\frac{k}{a^2}+\frac{8\pi G}{3}\rho, \quad
  \frac{\ddot{a}}{a}=\frac{\Lambda}{3}-\frac{4\pi G}{3}(\rho+3p),  \end{equation}
 where we note that $p<-\rho/3$ implies repulsive gravitation if $\Lambda=0$. Recaling the Hubble parameter $H=\dot{a}a^{-1}$  these equations can be written as
 \begin{equation}
  H^2=\frac{8\pi G}{3}\rho-\frac{k}{a^2}+\frac{\Lambda}{3}, \quad
  \dot{H}=-4\pi G(\rho+p)+\frac{k}{a^2}.
    \end{equation} 
 If the FRW  spacetime is filled with a fluid of energy density $\rho$ and pressure $p$, then the conservation law coud be derived from the Friedmann equations (2.5) as
 \begin{equation}
\dot{\rho}+3H(\rho+p)=0. \end{equation}
  In this work, we consider the case: $k=\Lambda=0$ and set $8\pi G=1$. 
 So finally the equations for the action (2.1) we can write in   \textit{the H-form}
	\begin{eqnarray}
		p&=&-2\dot{H}-3H^2,\\
	\rho&=&3H^2,\\
	\dot{\rho}&=&-3H(\rho+p),
	\end{eqnarray}
	 in  \textit{the N-form}	\begin{eqnarray}
		p&=&-2\ddot{N}-3\dot{N}^2,\\
	\rho&=&3\dot{N}^2,\\
	\dot{\rho}&=&-3\dot{N}(\rho+p)
	\end{eqnarray}
	or in   \textit{the a-form}	\begin{eqnarray}
		p&=&-2\frac{\ddot{a}}{a}-\frac{\dot{a}^2}{a^2},\\
	\rho&=&3\frac{\dot{a}^2}{a^2},\\
	\dot{\rho}&=&-\frac{3\dot{a}}{a}(\rho+p),
	\end{eqnarray}
	where $N=\ln{a}$.  Note that  one equation among three is redundant. Here we may choose the first two Friedmann equations as two relevant equations. To find $H$ we note that the general solution of the equation (2.7) we can write as
 \begin{equation}
  H(a)=\sqrt{-a^{-3}\int pa^2da}. 
 \end{equation}
 This formula tells us that  the EoS has the form
 \begin{equation}
  \rho=-3a^{-3}\int pa^2da 
 \end{equation}
 or
 \begin{equation}
  p=-\rho-3^{-1}a\rho_a 
 \end{equation}
 so that its parameter takes the form
 \begin{equation}
  \omega=-1-3^{-1}a(\ln \rho)_a. 
 \end{equation}
	
	\section{Differential equations of state: 0.5 - models}
\subsection{Integrable 0.5 - models}
In this subsection we consider some integrable 0.5 - models that means models with the half EoS ($half\equiv 0.5$).  As the integrable cells  (which we inplant to the body of the original gravitational  system of equations), we use   Painlev$\acute{e}$ equations. The function of these integrable cells are to convert the original gravitational (nonintegrable) system into the integrable system.
	\subsubsection{$\rho$ - models}

 Let's go further or,  if exactly, go back. To the Friedmann equations e.g. (2.7)-(2.9) and consider the following its extension
 \begin{eqnarray}
		p&=&-2\dot{H}-3H^2,\\
	\rho&=&3H^2,\\
\ddot{\rho}&=&\rho^{-1}\dot{\rho}^2-t^{-1}(\dot{\rho}-\alpha\rho^2-\beta)+\gamma\rho^3+\delta\rho^{-1},\\
	\dot{\rho}&=&-3H(\rho+p),
	\end{eqnarray}
 where $\alpha,\beta, \gamma, \delta$ are consts.  It is the $A_{IIIA}$ - model (see below). We guess that this system is integrable due to integrability of the equation (3.3) which is nothing but the P$_{III}$ - equation and which plays the role of the integrable cell (see e.g. \cite{Kuralay2}-\cite{Ablowitz}). To solve it, we start from the equation (3.3). In particular, it has the following well-known particular solutions \cite{Ablowitz}:
  \begin{eqnarray}
  \rho(t)&\equiv &\rho(t;\mu,-\mu\kappa^2,\lambda,-\lambda\kappa^4)=\kappa,\\
 \rho(t)&\equiv &\rho(t;0,-\mu,0,\mu\kappa)=\kappa t, \\ 
  \rho(t)&\equiv &\rho(t;2\kappa+3,-2\kappa+1,1,-1)=\frac{t+\kappa}{t+\kappa+1},\\
  \rho(t)&\equiv& \rho(t;\mu,0,0,-\mu\kappa^3)=\kappa \sqrt[3]{t},\\
  \rho(t)&\equiv& \rho(t;0,-2\kappa,0,4\kappa\mu-\lambda^2)=t[\kappa(\ln t)^2+\lambda\ln t+\mu],\\
\rho(t)&\equiv& \rho(t;-\nu^2\lambda,0,\nu^2(\lambda^2-4\kappa\mu),0)=\frac{t^{\nu-1}}{\kappa t^{2\nu}+\lambda t^{\nu}+\mu},\\
 \rho(t)&\equiv& \rho(t;0.5\epsilon)=-\epsilon_1(\ln{\phi})_{t}
 \end{eqnarray}
and so on. Here 
 \begin{equation}
  \phi(t)=t^{\nu}[C_1J_{\nu}(\zeta)+C_2Y_{\nu}(\zeta)],
 \end{equation}
 where $C_i=consts, \quad \zeta=\sqrt{\epsilon_1\epsilon_2}t,\quad \nu=0.5\alpha\epsilon_1$ and $J_{\nu}(\zeta), Y_{\nu}(\zeta)$ are Bessel functions. We are here modest and  work  just with the simplest  solutions. 
 
 i) First let's consider  the solution (3.5):
  \begin{equation}
  \rho(t)=\kappa=const. 
 \end{equation}
From  (3.2) we get that
 \begin{equation}
  H(t)=H_0=3^{-0.5}\kappa^{0.5}=const, 
 \end{equation}
so that this solution corresponds to the de Sitter case. 

ii) Next, let's consider the other simplest solution namely the solution (3.6):  \begin{equation}
  \rho(t)=\kappa t, 
 \end{equation}
  where we assume that $\kappa>0$. Then the equation (3.2) gives
 \begin{equation}
  H=\pm\sqrt{\frac{\kappa}{3}}t^{0.5}
 \end{equation}
 so that we have
  \begin{equation}
  a=a_0e^{\pm\sqrt{\frac{4\kappa}{27
  }}t^{1.5}}.
 \end{equation}
For this particular solution  the  EoF has the form
  \begin{equation}
  p=-\rho \mp 3^{-0.5}\kappa^{0.75}\rho^{-0.25}.
 \end{equation}
   The corresponding EoF parameter reads as
  \begin{equation}
  \omega=-1\mp 3^{-0.5}\kappa^{-0.5}t^{-1.5}.
 \end{equation}
 For this case we have
  \begin{equation}
  \dot{a}=\pm a_0\sqrt{\frac{\kappa}{3}}t^{0.5}e^{\pm\sqrt{\frac{4\kappa}{27
  }}t^{1.5}}
 \end{equation}
 and
  \begin{equation}
  \ddot{a}=\pm \frac{\kappa a_0}{3}t^{-0.5}[\sqrt{\frac{3}{4\kappa}}\pm t^{1.5}]e^{\pm\sqrt{\frac{4\kappa}{27
  }}t^{1.5}}.
 \end{equation}
 Let's calculate the deceleration parameter. We have
   \begin{equation}
  q=-t^{-1.5}[\sqrt{\frac{3}{4\kappa}}\pm t^{1.5}].
 \end{equation}
 Hence we see that $q=0$  as $t_0=\sqrt[3]{\frac{3}{4\kappa}}$. It means that  this solution describes the deceleration and  acceleration phases of the expansion of the  universe and $t_0=\sqrt[3]{\frac{3}{4\kappa}}$ is the transition point. Similarly, we can explore the other solutions of the $A_{IIIA}$ - model (3.1) - (3.4) as well as the other  $\rho$ - models. Now we present the list of  $\rho$ - models.
 \\
 1) {\bf A$_{I}$ - models.}
  \begin{equation}
 \begin{array}{|c|c|}
\hline \\
A_{IA} - model &  \ddot{\rho}=6\rho^{2}+t \\
\hline \\
A_{IB} - model &  \rho_{aa}=6\rho^2+a  \\
\hline \\
A_{IC} - model &  \rho_{NN}=6\rho^2+N  \\
\hline \\
A_{ID} - model &  \rho_{HH}=6\rho^2+H \\
\hline
\end{array} \end{equation}
\\
 2) {\bf A$_{II}$ - models.} \\ 
  \begin{equation} \begin{array}{|c| c|}
\hline \\
A_{IIA} - model &  \ddot{\rho}=2\rho^3+t\rho+\alpha   \\
\hline \\
A_{IIB} - model &  \rho_{aa}=2\rho^3+a\rho+\alpha  \\
\hline \\
A_{IIC} - model &  \rho_{NN}=2\rho^3+N\rho+\alpha   \\
\hline \\
A_{IID} - model &  \rho_{HH}=2\rho^3+H\rho+\alpha  \\
\hline
\end{array} \end{equation}
 3) {\bf $A_{III}$ - models.} \\
  \begin{equation}
 \begin{array}{|c|c|}
\hline \\
A_{IIIA} - model &  \ddot{\rho}=\rho^{-1}\dot{\rho}^2-t^{-1}(\dot{\rho}-\alpha \rho^2-\beta)+\gamma \rho^3+\delta \rho^{-1}  \\
\hline \\
A_{IIIB} - model &  \rho_{aa}=\rho^{-1}\rho_{a}^2-a^{-1}(\rho_{a}-\alpha \rho^2-\beta)+\gamma \rho^3+\delta \rho^{-1} \\
\hline \\
A_{IIIC} - model &  \rho_{NN}=\rho^{-1}\rho_{N}^2-N^{-1}(\rho_{N}-\alpha \rho^2-\beta)+\gamma \rho^3+\delta \rho^{-1}  \\
\hline \\
A_{IIID} - model &  \rho_{HH}=\rho^{-1}\rho_{H}^2-H^{-1}(\rho_{H}-\alpha \rho^2-\beta)+\gamma \rho^3+\delta \rho^{-1} \\
\hline
\end{array} \end{equation}
\vspace{1cm}\\
 4) {\bf $A_{IV}$ - models.} 
  \\
  \begin{equation}
 \begin{array}{|c|c|}
\hline \\
A_{IVA} - model &  \ddot{\rho}=0.5\rho^{-1}\dot{\rho}^2+1.5\rho^3+4t\rho^2+2(t^2-\alpha)\rho+\beta \rho^{-1}  \\
\hline \\
A_{IVB} - model &  \rho_{aa}=0.5\rho^{-1}\rho_{a}^2+1.5\rho^3+4a\rho^2+2(a^2-\alpha)\rho+\beta \rho^{-1}\\
\hline \\
A_{IVC} - model &  \rho_{NN}=0.5\rho^{-1}\rho_{N}^2+1.5\rho^3+4N\rho^2+2(N^2-\alpha)\rho+\beta \rho^{-1} \\
\hline \\
A_{IVD} - model &  \rho_{HH}=0.5\rho^{-1}\rho_{H}^2+1.5\rho^3+4H\rho^2+2(H^2-\alpha)\rho+\beta \rho^{-1}\\
\hline
\end{array} \end{equation}
\vspace{1cm}\\
 5) {\bf $A_{V}$ - models.} 
  \\
  \begin{equation}
 \begin{array}{|c|c|}
\hline \\
A_{VA} - model &  \ddot{\rho}=\phi\dot{\rho}^2-t^{-1}(\dot{\rho}-\gamma \rho)+t^{-2}(\rho-1)^2(\alpha \rho+\beta \rho^{-1})+\delta \psi  \\
\hline \\
A_{VB} - model &  \rho_{aa}=\phi \rho_{a}^2-a^{-1}(\rho_{a}-\gamma \rho)+a^{-2}(\rho-1)^2(\alpha \rho+\beta \rho^{-1})+\delta \psi  \\
\hline \\
A_{VC} - model &  \rho_{NN}=\phi \rho_{N}^2-N^{-1}(\rho_{N}-\gamma \rho)+N^{-2}(\rho-1)^2(\alpha \rho+\beta \rho^{-1})+\delta \psi   \\
\hline \\
A_{VD} - model &  \rho_{HH}=\phi \rho_{H}^2-H^{-1}(\rho_{H}-\gamma \rho)+H^{-2}(\rho-1)^2(\alpha \rho+\beta \rho^{-1})+\delta \psi   \\
\hline
\end{array} \end{equation}
Here $\phi=0.5\rho^{-1}+(\rho-1)^{-1}, \quad \psi= \rho(\rho+1)(\rho-1)^{-1}$.
\vspace{1cm}\\
 6) {\bf $A_{VI}$ - models.}
  \\
  \begin{equation}
 \begin{array}{|c|c|}
\hline \\
A_{VIA} - model &  \ddot{\rho}=\varphi(t)\dot{\rho}^2-\xi(t)\dot{\rho}-\frac{\rho(\rho-1)(\rho-t)}{t^{2}(t-1)^{2}}[\alpha +\frac{\beta t}{ \rho^{2}}+\zeta(t)]  \\
\hline \\
A_{VIB} - model &  \rho_{aa}=\varphi(a)\rho_{a}^2-\xi(a) \rho_{a}-\frac{\rho(\rho-1)(\rho-a)}{a^{2}(a-1)^{2}}[\alpha +\frac{\beta a}{ \rho^{2}}+\zeta(a)]  \\
\hline \\
A_{VIC} - model &  \rho_{NN}=\varphi(N) \rho_{N}^2-\xi(N) \rho_{N}-\frac{\rho(\rho-1)(\rho-N)}{N^{2}(N-1)^{2}}[\alpha +\frac{\beta N}{ \rho^{2}}+\zeta(N)] \\
\hline \\
A_{VID} - model &  \rho_{HH}=\varphi(H) \rho_{H}^2-\xi(H) \rho_{H}-\frac{\rho(\rho-1)(\rho-H)}{H^{2}(H-1)^{2}}[\alpha +\frac{\beta H}{ \rho^{2}}+\zeta(H)] \\
\hline
\end{array} \end{equation}
Here $\varphi(z)=0.5[\rho^{-1}+(\rho-1)^{-1}+(\rho-z)^{-1}], \quad \xi(z)= [z^{-1}+(z-1)^{-1}+(\rho-z)^{-1}],\quad \zeta(z)=[\gamma(z-1)(\rho-1)^{-2}+\delta z(z-1)(\rho-z)^{-2}]$.

 	\subsubsection{p - models}
 	Now we want consider one of important part of integrable 0.5 models namely  p - models. Let's implant the integrable cell, e.g. the $B_{IIIB}$ - model (see below),  to the body of the gravitional equations.  The aim of this implantation is  the original (in general nonintegrable) system convert  into the integrable system.  If this integrable  cell is,  e.g. the $B_{IIIB}$ - model, as result  we get the following closed system of the equations
 	\begin{eqnarray}
		p&=&-2\dot{H}-3H^2,\\
	\rho&=&3H^2,\\
p_{aa}&=&p^{-1}p_{a}^2-a^{-1}(p_{a}-\alpha p^2-\beta)+\gamma p^3+\delta p^{-1},\\
	\dot{\rho}&=&-3H(\rho+p),
	\end{eqnarray}
 where $\alpha,\beta, \gamma, \delta$ are consts.  This system we also call  the $B_{IIIB}$ - model,  which is (as we expect and believe) integrable. Also we note that the equation (3.31) is the P$_{III}$ - equation and plays the role of  the integrable cell.  As an integrable, the system (3.29)-(3.32) admits  (may be infinite number)  exact solutions. The construction its exact solutions we start from the  equation (3.31).  In particular, the  equation (3.31) has the following particular solutions (see e.g. \cite{Ablowitz}):
  \begin{eqnarray}
  p(a)&\equiv &p(a;\mu,-\mu\kappa^2,\lambda,-\lambda\kappa^4)=\kappa,\\
 p(a)&\equiv &p(a;0,-\mu,0,\mu\kappa)=\kappa a, \\ 
  p(a)&\equiv &p(a;2\kappa+3,-2\kappa+1,1,-1)=\frac{a+\kappa}{a+\kappa+1},\\
  p(a)&\equiv& p(a;\mu,0,0,-\mu\kappa^3)=\kappa \sqrt[3]{a},\\
  p(a)&\equiv& p(a;0,-2\kappa,0,4\kappa\mu-\lambda^2)=a[\kappa(\ln a)^2+\lambda\ln a+\mu],\\
p(a)&\equiv& p(a;-\nu^2\lambda,0,\nu^2(\lambda^2-4\kappa\mu),0)=\frac{a^{\nu-1}}{\kappa a^{2\nu}+\lambda a^{\nu}+\mu},\\
 p(a)&\equiv& p(a;0.5\epsilon)=-\epsilon_1(\ln{\phi})_{a}
 \end{eqnarray}
and so on. Here 
 \begin{equation}
  \phi(a)=a^{\nu}[C_1J_{\nu}(\zeta)+C_2Y_{\nu}(\zeta)],
 \end{equation}
 where $C_i=consts, \quad \zeta=\sqrt{\epsilon_1\epsilon_2}a,\quad \nu=0.5\alpha\epsilon_1$ and $J_{\nu}(\zeta), Y_{\nu}(\zeta)$ are Bessel functions. Now we   consider some  simplest  solutions. 
 
 i) Let's start from   the solution (3.33)
 \begin{equation}
  p(a)=\kappa=const. 
 \end{equation}
 To find $H$ we use the formula  (2.16). As result  we get
 \begin{equation}
  H(a)=\sqrt{-a^{-3}(3^{-1}\kappa a^3+C)}=\sqrt{-3^{-1}\kappa -Ca^{-3}}=\sqrt{3^{-1}\rho_0  a^{-3}+3^{-1}\Lambda}. 
 \end{equation}
where $\rho_0=-3C, \quad \Lambda=-\kappa$.  It  is nothing but $\Lambda$CDM model. So this solution corresponds to  the  $\Lambda$CDM cosmology. In this case, the above formulas give
 \begin{equation}
  p=-\Lambda, \quad \rho=\rho_0  a^{-3}+\Lambda 
 \end{equation}
which corresponds the EoF parameter
\begin{equation}
 \omega=-1+\rho_0(\rho_0  +\Lambda a^{3})^{-1}. 
 \end{equation}
 
ii) Now let's we consider the next simplest solution namely the solution (3.34):\begin{equation}
  p(a)=\kappa a. 
 \end{equation}
 Eq. (2.16) gives
  \begin{equation}
  H(a)=\sqrt{-a^{-3}(0.25\kappa a^4+C)}. 
 \end{equation}
 The corresponding energy density is given by
   \begin{equation}
  \rho=-3a^{-3}(0.25\kappa a^4+C). 
 \end{equation}
 
For this particular solution  the  EoF and its parameter take  the form
   \begin{equation}
  \rho=-0.75p-3C\kappa^3p^{-3} 
 \end{equation}
and 
  \begin{equation}
  \omega=-\frac{\kappa a^4}{0.75 \kappa a^{4}+3C},
 \end{equation}
 respectively. In the limit $a\rightarrow \infty$ we get $\omega\rightarrow -4/3$.

Similarly we can write  $B_{J}$ - models induced by the other P$_{J}$ - equations. Here the list of such models.
 \\
1) {\bf B$_{I}$ - models.}
  \begin{equation}
 \begin{array}{|c|c|}
\hline \\
B_{IA} - model &  \ddot{p}=6p^{2}+t \\
\hline \\
B_{IB} - model &  p_{aa}=6p^2+a  \\
\hline \\
B_{IC} - model &  p_{NN}=6p^2+N  \\
\hline \\
B_{ID} - model &  p_{HH}=6p^2+H \\
\hline
\end{array} \end{equation}
\\
 2) {\bf B$_{II}$ - models.} \\ 
  \begin{equation} \begin{array}{|c| c|}
\hline \\
B_{IIA} - model &  \ddot{p}=2p^3+tp+\alpha   \\
\hline \\
B_{IIB} - model &  p_{aa}=2p^3+ap+\alpha  \\
\hline \\
B_{IIC} - model &  p_{NN}=2p^3+Np+\alpha   \\
\hline \\
B_{IID} - model &  p_{HH}=2p^3+Hp+\alpha  \\
\hline
\end{array} \end{equation}
 3) {\bf $B_{III}$ - models.} \\
  \begin{equation}
 \begin{array}{|c|c|}
\hline \\
B_{IIIA} - model &  \ddot{p}=p^{-1}\dot{p}^2-t^{-1}(\dot{p}-\alpha p^2-\beta)+\gamma p^3+\delta p^{-1}  \\
\hline \\
B_{IIIB} - model &  p_{aa}=p^{-1}p_{a}^2-a^{-1}(p_{a}-\alpha p^2-\beta)+\gamma p^3+\delta p^{-1} \\
\hline \\
B_{IIIC} - model &  p_{NN}=p^{-1}p_{N}^2-N^{-1}(p_{N}-\alpha p^2-\beta)+\gamma p^3+\delta p^{-1}  \\
\hline \\
B_{IIID} - model &  p_{HH}=p^{-1}p_{H}^2-H^{-1}(p_{H}-\alpha p^2-\beta)+\gamma p^3+\delta p^{-1} \\
\hline
\end{array} \end{equation}
\vspace{1cm}\\
 4) {\bf $B_{IV}$ - models.} 
  \\
  \begin{equation}
 \begin{array}{|c|c|}
\hline \\
B_{IVA} - model &  \ddot{p}=0.5p^{-1}\dot{p}^2+1.5p^3+4tp^2+2(t^2-\alpha)p+\beta p^{-1}  \\
\hline \\
B_{IVB} - model &  p_{aa}=0.5p^{-1}p_{a}^2+1.5p^3+4ap^2+2(a^2-\alpha)p+\beta p^{-1}\\
\hline \\
B_{IVC} - model &  p_{NN}=0.5p^{-1}p_{N}^2+1.5p^3+4Np^2+2(N^2-\alpha)p+\beta p^{-1} \\
\hline \\
B_{IVD} - model &  p_{HH}=0.5p^{-1}p_{H}^2+1.5p^3+4Hp^2+2(H^2-\alpha)p+\beta p^{-1}\\
\hline
\end{array} \end{equation}
\vspace{1cm}\\
 5) {\bf $B_{V}$ - models.} 
  \\
  \begin{equation}
 \begin{array}{|c|c|}
\hline \\
B_{VA} - model &  \ddot{p}=\phi\dot{p}^2-t^{-1}(\dot{p}-\gamma p)+t^{-2}(p-1)^2(\alpha p+\beta p^{-1})+\delta \psi  \\
\hline \\
B_{VB} - model &  p_{aa}=\phi p_{a}^2-a^{-1}(p_{a}-\gamma p)+a^{-2}(p-1)^2(\alpha p+\beta p^{-1})+\delta \psi  \\
\hline \\
B_{VC} - model &  p_{NN}=\phi p_{N}^2-N^{-1}(p_{N}-\gamma p)+N^{-2}(p-1)^2(\alpha p+\beta p^{-1})+\delta \psi   \\
\hline \\
B_{VD} - model &  p_{HH}=\phi p_{H}^2-H^{-1}(p_{H}-\gamma p)+H^{-2}(p-1)^2(\alpha p+\beta p^{-1})+\delta \psi   \\
\hline
\end{array} \end{equation}
Here $\phi=0.5p^{-1}+(p-1)^{-1}, \quad \psi= p(p+1)(p-1)^{-1}$.
\vspace{1cm}\\
 6) {\bf $B_{VI}$ - models.}
  \\
  \begin{equation}
 \begin{array}{|c|c|}
\hline \\
B_{VIA} - model &  \ddot{p}=\varphi(t)\dot{p}^2-\xi(t)\dot{p}-\frac{p(p-1)(p-t)}{t^{2}(t-1)^{2}}[\alpha +\frac{\beta t}{ p^{2}}+\zeta(t)]  \\
\hline \\
B_{VIB} - model &  p_{aa}=\varphi(a)p_{a}^2-\xi(a) p_{a}-\frac{p(p-1)(p-a)}{a^{2}(a-1)^{2}}[\alpha +\frac{\beta a}{ p^{2}}+\zeta(a)]  \\
\hline \\
B_{VIC} - model &  p_{NN}=\varphi(N) p_{N}^2-\xi(N) p_{N}-\frac{p(p-1)(p-N)}{N^{2}(N-1)^{2}}[\alpha +\frac{\beta N}{ p^{2}}+\zeta(N)] \\
\hline \\
B_{VID} - model &  p_{HH}=\varphi(H) p_{H}^2-\xi(H) p_{H}-\frac{p(p-1)(p-H)}{H^{2}(H-1)^{2}}[\alpha +\frac{\beta H}{ p^{2}}+\zeta(H)] \\
\hline
\end{array} \end{equation}
Here $\varphi(z)=0.5[p^{-1}+(p-1)^{-1}+(p-z)^{-1}], \quad \xi(z)= [z^{-1}+(z-1)^{-1}+(p-z)^{-1}],\quad \zeta(z)=[\gamma(z-1)(p-1)^{-2}+\delta z(z-1)(p-z)^{-2}]$.
 
\subsection{Nonintegrable 0.5 - models}
Now we give some examples nonintegrable 0.5 - models.
\subsubsection{$\rho$ - models}
1) {\bf $\Lambda$ cosmology}.\\
i) Example 1:
\begin{equation}
 \ddot{\rho}=0.5\Lambda-1.5\dot{\rho}^2.\end{equation}
 ii)   Example 2:
\begin{equation}
 \rho_{aa}=0.5\Lambda-1.5\rho_{a}^2.\end{equation}
 iii)  Example 3:
\begin{equation}
 \rho_{HH}=0.5\Lambda-1.5\rho_{H}^2.\end{equation}
   iv)   Example 4:
\begin{equation}
 \rho_{NN}=0.5\Lambda-1.5\rho_{N}^2.\end{equation}
   2) {\bf Pinney  cosmology}.  It induced by  the Pinney equation. Let us present 4 examples of such models.
 \\
 i)  Example 1:
 \begin{equation}
 \ddot{\rho}=\xi(t)\rho+\frac{k}{\rho^3}.\end{equation}
 \\
ii)  Example 2:
 \begin{equation}
 \rho_{aa}=\xi(a)\rho+\frac{k}{\rho^3}.\end{equation}
  \\
iii)  Example 3:
 \begin{equation}
 \rho_{HH}=\xi(H)\rho+\frac{k}{\rho^3}.\end{equation}
   \\
iv)  Example 4:
 \begin{equation}
 \rho_{NN}=\xi(N)\rho+\frac{k}{\rho^3}.\end{equation}
  3) {\bf Schr$\ddot{o}$dinger cosmology}.  For this model we give 4 submodels. 
\\
i) Example 1:
 \begin{equation}
 \ddot{\rho}=u(t)\rho+k\rho,\end{equation}
 \\
ii) Example 2:
 \begin{equation}
 \rho_{aa}=u(a)\rho+k\rho.\end{equation}
\\
iii) Example 3:
 \begin{equation}
 \rho_{HH}=u(H)\rho+k\rho.\end{equation}
 \\
iv) Example 4:
 \begin{equation}
 \rho_{NN}=u(N)\rho+k\rho.\end{equation}
  4) {\bf Hypergeometric cosmology}. Let us we present  4 examples.
  \\ 
  i) Example 1:
  \begin{equation}
 \ddot{\rho}=t^{-1}(1-t)^{-1}\{[(\alpha+\beta+1)t-\gamma]\dot{\rho}+\alpha\beta \rho\}.\end{equation}
 \\
 ii) Example 2:
  \begin{equation}
 \rho_{aa}=a^{-1}(1-a)^{-1}\{[(\alpha+\beta+1)a-\gamma]\rho_{a}+\alpha\beta \rho\}.\end{equation}
 \\
  iii) Example 3:
  \begin{equation}
 \rho_{HH}=H^{-1}(1-H)^{-1}\{[(\alpha+\beta+1)H-\gamma]\rho_{H}+\alpha\beta \rho\}.\end{equation}
\\
  iv) Example 4:
  \begin{equation}
 \rho_{NN}=N^{-1}(1-N)^{-1}\{[(\alpha+\beta+1)N-\gamma]\rho_{N}+\alpha\beta \rho\}.\end{equation}

\subsubsection{$p$ - models}
1) {\bf $\Lambda$ cosmology}.\\
i) Example 1:
\begin{equation}
 \ddot{p}=0.5\Lambda-1.5\dot{p}^2.\end{equation}
 ii)   Example 2:
\begin{equation}
 p_{aa}=0.5\Lambda-1.5p_{a}^2.\end{equation}
 iii)  Example 3:
\begin{equation}
 p_{HH}=0.5\Lambda-1.5p_{H}^2.\end{equation}
   iv)   Example 4:
\begin{equation}
 p_{NN}=0.5\Lambda-1.5p_{N}^2.\end{equation}
   2) {\bf Pinney  cosmology}.  It induced by  the Pinney equation. Let us present 4 examples of such models.
 \\
 i)  Example 1:
 \begin{equation}
 \ddot{p}=\xi(t)p+\frac{k}{p^3},\end{equation}
 where $\xi=\xi(t),\quad k=const$.
 \\
ii)  Example 2:
 \begin{equation}
 p_{aa}=\xi(a)p+\frac{k}{p^3}.\end{equation}
  \\
iii)  Example 3:
 \begin{equation}
 p_{HH}=\xi(H)p+\frac{k}{p^3}.\end{equation}
   \\
iv)  Example 4:
 \begin{equation}
 p_{NN}=\xi(N)p+\frac{k}{p^3}.\end{equation}
  3) {\bf Schr$\ddot{o}$dinger cosmology}.  For this model also write 5 submodels. 
\\
i) Example 1:
 \begin{equation}
 \ddot{p}=u(t)p+kp,\end{equation}
 where $u=u(t),\quad k=const$. 
 \\
ii) Example 2:
 \begin{equation}
 p_{aa}=u(a)p+kp.\end{equation}
\\
iii) Example 3:
 \begin{equation}
 p_{HH}=u(H)p+kp.\end{equation}
 \\
iv) Example 4:
 \begin{equation}
 p_{NN}=u(N)p+kp.\end{equation}
  4) {\bf Hypergeometric cosmology}. Let us we present  4 examples.
  \\ 
  i) Example 1:
  \begin{equation}
 \ddot{p}=t^{-1}(1-t)^{-1}\{[(\alpha+\beta+1)t-\gamma]\dot{p}+\alpha\beta p\}.\end{equation}
 \\
 ii) Example 2:
  \begin{equation}
 p_{aa}=a^{-1}(1-a)^{-1}\{[(\alpha+\beta+1)a-\gamma]p_{a}+\alpha\beta p\}.\end{equation}
 \\
  iii) Example 3:
  \begin{equation}
 p_{HH}=H^{-1}(1-H)^{-1}\{[(\alpha+\beta+1)H-\gamma]p_{H}+\alpha\beta p\}.\end{equation}
\\
  iv) Example 4:
  \begin{equation}
 p_{NN}=N^{-1}(1-N)^{-1}\{[(\alpha+\beta+1)N-\gamma]p_{N}+\alpha\beta p\}.\end{equation}
\section{Differential equations of state: 1.0 - models}
Here we present some examples integrable and nonintegrable  1.0 - models that means  models with the full EoS ($full\equiv 1.0$).
\subsection{Integrable 1.0 - models}
Here the list integrable 1.0 - models.
  \begin{equation}
 \begin{array}{|c|c|}
\hline \\
K_{I} - model &   p_{\rho\rho}=6p^2+\rho  \\
\hline \\
K_{II} - model &  p_{\rho\rho}=2p^3+\rho p+\alpha  \\
\hline \\
K_{III} - model &  p_{\rho\rho}=\frac{1}{p}p_\rho^2-\frac{1}{\rho}(p_\rho-\alpha p^2-\beta)+\gamma p^3+\frac{\delta}{p} \\
\hline \\
K_{IV} - model &  p_{\rho\rho}=\frac{1}{2p}p_\rho^2+1.5p^3+4\rho p^2+2(\rho^2-\alpha)p+
 \frac{\delta}{p} \\
\hline \\
K_{V} - model &  p_{\rho\rho}=(\frac{1}{2p}+\frac{1}{p-1})p_\rho^2
 -\frac{1}{\rho}(p_\rho-\gamma  p)+\rho^{-2}(p-1)^2(\alpha p+\beta p^{-1})
 +\frac{\delta p(p+1)}{p-1} \\
\hline \\
K_{VI} - model &  p_{\rho\rho}=\varphi(\rho)p_\rho^2
 -\xi(\rho)p_\rho
 +\rho^{-2}(\rho-1)^{-2}p(p-1)(p-\rho)[\alpha+ \beta \rho p^{-2}+\zeta(\rho)]\\
\hline
\end{array} \end{equation}
Here $\varphi(z)=0.5[p^{-1}+(p-1)^{-1}+(p-z)^{-1}], \quad \xi(z)= [z^{-1}+(z-1)^{-1}+(p-z)^{-1}],\quad \zeta(z)=[\gamma(z-1)(p-1)^{-2}+\delta z(z-1)(p-z)^{-2}]$. All above presented $K_{J}$ - models admit exact solutions. Now let's present some of these solutions.
	\subsubsection{K$_{I}$ - model}
\begin{equation}
 p_{\rho\rho}=6p^2+\rho.\end{equation}
\subsubsection{K$_{II}$ - model}
\begin{equation}
 p_{\rho\rho}=2p^3+\rho p+\alpha.\end{equation}
 This models has the following particular solutions 
  \begin{eqnarray}
  p&\equiv &p(\rho;1.5)=\psi-(2\psi^2+\rho)^{-1},\\
 p&\equiv &p(\rho;1)=-\frac{1}{\rho}, \\ 
  p&\equiv &p(\rho;2)=\frac{1}{\rho}-\frac{3\rho^2}{\rho^3+4},\\
  p&\equiv& p(\rho;3)=\frac{3\rho^2}{\rho^3+4}-\frac{6\rho^2(\rho^3+10)}{\rho^6+20\rho^3-80},\\
p&\equiv& p(\rho;4)=-\frac{1}{\rho}+\frac{6\rho^2(\rho^3+10)}{\rho^6+20\rho^3-80}-\frac{9\rho^5(\rho^3+40)}{\rho^9+60\rho^6+11200},\\
 p&\equiv& p(\rho;0.5\epsilon)=-\epsilon\psi
 \end{eqnarray}
and so on. Here 
 \begin{equation}
\psi=(\ln{\phi})_{\rho}, \quad  \phi(\rho)=C_1Ai(-2^{-1/3}\rho)+C_2Bi{(-2^{-1/3}\rho)},\end{equation} 
where
 $C_i=consts$ and $Ai(x), Bi(x)$ are Airy functions.

 \subsubsection{K$_{III}$ - model}
\begin{equation}
 p_{\rho\rho}=\frac{1}{p}p_\rho^2-\frac{1}{\rho}(p_\rho-\alpha p^2-\beta)+\gamma p^3+\frac{\delta}{p}.\end{equation}
This equation admits the infinite number exact solutions. For example, it has the following particular solutions (see e.g. \cite{Ablowitz}) \begin{eqnarray}
  p&\equiv &p(\rho;\mu,-\mu\kappa^2,\lambda,-\lambda\kappa^4)=\kappa,\\
 p&\equiv &p(\rho;0,-\mu,0,\mu\kappa)=\kappa \rho, \\ 
  p&\equiv &p(\rho;2\kappa+3,-2\kappa+1,1,-1)=\frac{\rho+\kappa}{\rho+\kappa+1},\\
  p&\equiv& p(\rho;\mu,0,0,-\mu\kappa^3)=\kappa \sqrt[3]{\rho},\\
  p&\equiv& p(\rho;0,-2\kappa,0,4\kappa\mu-\lambda^2)=\rho[\kappa(\ln \rho)^2+\lambda\ln \rho+\mu],\\
p&\equiv& p(\rho;-\nu^2\lambda,0,\nu^2(\lambda^2-4\kappa\mu),0)=\frac{\rho^{\nu-1}}{\kappa \rho^{2\nu}+\lambda \rho^{\nu}+\mu},\\
 p&\equiv& p(\rho;0.5\epsilon)=-\epsilon_1(\ln{\phi})_{\rho}
 \end{eqnarray}
and so on. Here 
 \begin{equation}
  \phi(\rho)=\rho^{\nu}[C_1J_{\nu}(\zeta)+C_2Y_{\nu}(\zeta)],
 \end{equation}
 where $C_i=consts, \quad \zeta=\sqrt{\epsilon_1\epsilon_2}\rho,\quad \nu=0.5\alpha\epsilon_1$ and $J_{\nu}(\zeta), Y_{\nu}(\zeta)$ are Bessel functions.  
 
 \subsubsection{K$_{IV}$ - model}
\begin{equation}
 p_{\rho\rho}=\frac{1}{2p}p_\rho^2+1.5p^3+4\rho p^2+2(\rho^2-\alpha)p+
 \frac{\delta}{p}.\end{equation}
 This equation  has the following particular solutions (see e.g. \cite{Ablowitz}): \begin{eqnarray}
  p&\equiv &p(\rho;\pm 2,-2)=\pm \rho^{-1},\\
 p&\equiv &p(\rho;0,-2)=-2\rho, \\ 
  p&\equiv &p(\rho;0,-3^{-2}2)=-3^{-1}2\rho,\\
  p&\equiv& p(\rho;-m,-2(m-1)^2)=-[\ln H_{m-1}(\rho)]_{\rho},\\
  p&\equiv&p(\rho) =2i\pi^{-0.5}e^{ \rho^2}[iC+ erfc(i\rho)]^{-1},\\
p&\equiv&p(\rho) =2\pi^{-0.5}e^{ -\rho^2}[C- erfc(\rho)]^{-1},\\
 p&\equiv& p(\rho;0.5\epsilon)=-\epsilon_1(\ln{\phi})_{\rho}
 \end{eqnarray}
and so on. Here 
 \begin{equation}
  \phi(\rho)=[C_1U(\zeta, 2^{0.5}\rho)+C_2V(\zeta, 2^{0.5}\rho)]e^{0.5\epsilon \rho^2},
 \end{equation}
 where $C_i=consts, \quad \zeta=\alpha+0.5\epsilon,\quad \nu=0.5\alpha\epsilon_1$ and $U, V)$ are parabolic cylinder  functions, $H_{m}$ are Hermite polynomials, $erfc$ is the complementary error function (for detail see e.g. \cite{Ablowitz}).  
 
  \subsubsection{K$_{V}$ - model}
\begin{equation}
 p_{\rho\rho}=(\frac{1}{2p}+\frac{1}{p-1})p_\rho^2
 -\frac{1}{\rho}(p_\rho-\gamma  p)+\rho^{-2}(p-1)^2(\alpha p+\beta p^{-1})
 +\frac{\delta p(p+1)}{p-1}.\end{equation}
 This equation admits  the following particular solutions (see e.g. \cite{Ablowitz}): \begin{eqnarray}
  p&\equiv &p(\rho;0.5,-0.5\mu^2,\kappa(2-\mu),-0.5\kappa^2)=\kappa  \rho+\mu,\\
 p&\equiv &p(\rho;0.5,\kappa^2\mu,2\kappa\mu,\mu)=\kappa ( \rho+\kappa)^{-1}, \\ 
  p&\equiv &p(\rho;0.125,-0.125,-\kappa\mu,\mu)=(\kappa+  \rho)(\kappa-\rho)^{-1},\\
  p&\equiv& p(\rho;\mu,-0.125, -\mu\kappa^2,0)=1+\kappa  \rho^{0.5},\\
  p&\equiv&p(\rho;0,0,\mu,-0.5\mu^2)=\kappa e^{\mu \rho},\\
 p&\equiv& p(\rho)=  -\epsilon_1\rho(\ln{\phi})_{\rho}
 \end{eqnarray}
and so on. Here 
 \begin{equation}
\phi(\rho)=(\epsilon_2\rho)^{-\nu}[C_1M_{\kappa,\mu}(\epsilon_2\rho)+C_2W_{\kappa,\mu}(\epsilon_2\rho)]e^{0.5\epsilon_2\rho},
 \end{equation}
 where $C_i=consts$ and $M, W$ are Whittaker  functions (for more exact details see e.g. \cite{Ablowitz}).  
 
  \subsubsection{K$_{VI}$ - model}
$$
 p_{\rho\rho}=0.5\left(\frac{1}{p}+\frac{1}{p-1}+\frac{1}{p-\rho}\right)p_\rho^2
 -\left(\frac{1}{\rho}+\frac{1}{\rho-1}+\frac{1}{p-\rho}\right)p_\rho
 $$\begin{equation}+\rho^{-2}(\rho-1)^{-2}p(p-1)(p-\rho)\left[\alpha+ \beta \rho p^{-2}+\gamma(\rho-1)(p-1)^{-2}+\delta \rho(\rho-1)(p-\rho)^{-2}\right].\end{equation}
 This equation is integrable and  has the following particular solutions (see e.g. \cite{Ablowitz}): \begin{eqnarray}
   p&\equiv &p(\rho;\mu,-\mu\kappa^2,0.5,0.5-\mu(\kappa-1)^2)=\kappa  \rho,\\
 p&\equiv &p(\rho;0,0,2,0)=\kappa\rho^{2}, \\ 
  p&\equiv &p(\rho;0,0,0.5,-1.5)=\kappa\rho^{-1},\\
  p&\equiv& p(\rho;0,0,2,-4)=\kappa  \rho^{-2},\\
  p&\equiv&p(\rho;0.5(\kappa+\mu)^2,-0.5,0.5(\mu-1)^2,0.5\kappa(2-\kappa))=\rho(\kappa+
  \mu \rho)^{-1},\\
   p&\equiv&p(\rho;0.5\kappa^2,-0.5\kappa^2,0.5\mu^2,0.5(1-\mu^2))=\rho^{0.5}
 \end{eqnarray}
and so on (for more exact details see e.g. \cite{Ablowitz}).

\subsection{Nonintegrable 1.0 - models}

1) Example 1:
\begin{equation}
 p_{\rho\rho}=0.5\Lambda-1.5p_{\rho}^2.\end{equation}
  2)  Example 2:
 \begin{equation}
 p_{\rho\rho}=\xi(\rho)p+\frac{k}{p^3},\end{equation}
 3)  Example 3:
 \begin{equation}
 p_{\rho\rho}=u(\rho)p+kp,\end{equation}
  4)  Example 4:
  \begin{equation}
p_{\rho\rho}=\rho^{-1}(1-\rho)^{-1}\{[(\alpha+\beta+1)\rho-\gamma]\dot{p}+\alpha\beta p\}.\end{equation}
 5)  Example 5:
\begin{equation}
 p_{\rho\rho}=n(n-1)\sqrt[n]{A^2p^{n-2}}.
 \end{equation}
 This model  has following solution
 \begin{equation}
 p=-A\rho^{n}.
 \end{equation}
 	\section{D - models}
 	In this section we would like to consider the so-called \textit{D - models}, where  $D=D(p,\rho)$ is an arbitrary function of $\rho$ and $p$. Here some examples of such models.  
 	\subsection{Integrable D - models}
 1) {\bf  $D_{I}$ - models.}
 \begin{equation}
 \begin{array}{|c|c|}
\hline \\
D_{IA} - model &  \ddot{D}=6D^2+t   \\
\hline \\
D_{IB} - model &  D_{aa}=6D^2+a \\
\hline \\
D_{IC} - model &  D_{NN}=6D^2+N  \\
\hline \\
D_{ID} - model &  D_{HH}=6D^2+H \\
\hline \\
D_{IE} - model &  D_{\rho\rho}=6D^2+\rho\\
\hline \\
D_{IF} - model &  D_{pp}=6D^2+p \\
\hline
\end{array} \end{equation}
\vspace{1cm}\\
 2) {\bf $D_{II}$ - models.}\\ 
  \begin{equation} \begin{array}{|c| c|}
\hline \\
D_{IIA} - model &  \ddot{D}=2D^3+tD+\alpha   \\
\hline \\
D_{IIB} - model &  D_{aa}=2D^3+aD+\alpha  \\
\hline \\
D_{IIC} - model &  D_{NN}=2D^3+ND+\alpha   \\
\hline \\
D_{IID} - model &  D_{HH}=2D^3+HD+\alpha  \\
\hline \\
D_{IIE} - model &  D_{\rho\rho}=2D^3+\rho D+\alpha \\
\hline \\
D_{IIF} - model &  D_{pp}=2D^3+pD+\alpha  \\
\hline
\end{array} \end{equation}
 3) {\bf $D_{III}$ - models.} \\
  \begin{equation}
 \begin{array}{|c|c|}
\hline \\
D_{IIIA} - model &  \ddot{D}=D^{-1}\dot{D}^2-t^{-1}(\dot{D}-\alpha D^2-\beta)+\gamma D^3+\delta D^{-1}  \\
\hline \\
D_{IIIB} - model &  D_{aa}=D^{-1}D_{a}^2-a^{-1}(D_{a}-\alpha D^2-\beta)+\gamma D^3+\delta D^{-1} \\
\hline \\
D_{IIIC} - model &  D_{NN}=D^{-1}D_{N}^2-N^{-1}(D_{N}-\alpha D^2-\beta)+\gamma D^3+\delta D^{-1}  \\
\hline \\
D_{IIID} - model &  D_{HH}=D^{-1}D_{H}^2-H^{-1}(D_{H}-\alpha D^2-\beta)+\gamma D^3+\delta D^{-1} \\
\hline \\
D_{IIIE} - model &  D_{\rho\rho}=D^{-1}D_{\rho}^2-\rho^{-1}(D_{\rho}-\alpha D^2-\beta)+\gamma D^3+\delta D^{-1}\\
\hline \\
D_{IIIF} - model &  D_{pp}=D^{-1}D_{p}^2-p^{-1}(D_{p}-\alpha D^2-\beta)+\gamma D^3+\delta D^{-1}\\
\hline
\end{array} \end{equation}
\vspace{1cm}\\
 4) {\bf $D_{IV}$ - models.} 
  \\
  \begin{equation}
 \begin{array}{|c|c|}
\hline \\
D_{IVA} - model &  \ddot{D}=0.5D^{-1}\dot{D}^2+1.5D^3+4tD^2+2(t^2-\alpha)D+\beta D^{-1}  \\
\hline \\
D_{IVB} - model &  D_{aa}=0.5D^{-1}D_{a}^2+1.5D^3+4aD^2+2(a^2-\alpha)D+\beta D^{-1}\\
\hline \\
D_{IVC} - model &  D_{NN}=0.5D^{-1}D_{N}^2+1.5D^3+4ND^2+2(N^2-\alpha)D+\beta D^{-1} \\
\hline \\
D_{IVD} - model &  D_{HH}=0.5D^{-1}D_{H}^2+1.5D^3+4HD^2+2(H^2-\alpha)D+\beta D^{-1}\\
\hline \\
D_{IVE} - model &  D_{\rho\rho}=0.5D^{-1}D_{\rho}^2+1.5D^3+4\rho D^2+2(\rho^2-\alpha)D+\beta D^{-1}\\
\hline \\
D_{IVF} - model &  D_{pp}=0.5D^{-1}D_{p}^2+1.5D^3+4pD^2+2(p^2-\alpha)D+\beta D^{-1}\\
\hline
\end{array} \end{equation}
\vspace{1cm}\\
 5) {\bf $D_{V}$ - models.} 
  \\
  \begin{equation}
 \begin{array}{|c|c|}
\hline \\
D_{VA} - model &  \ddot{D}=\phi\dot{D}^2-t^{-1}(\dot{D}-\gamma D)+t^{-2}(D-1)^2(\alpha D+\beta D^{-1})+\delta \psi  \\
\hline \\
D_{VB} - model &  D_{aa}=\phi D_{a}^2-a^{-1}(D_{a}-\gamma D)+a^{-2}(D-1)^2(\alpha D+\beta D^{-1})+\delta \psi  \\
\hline \\
D_{VC} - model &  D_{NN}=\phi D_{N}^2-N^{-1}(D_{N}-\gamma D)+N^{-2}(D-1)^2(\alpha D+\beta D^{-1})+\delta \psi   \\
\hline \\
D_{VD} - model &  D_{HH}=\phi D_{H}^2-H^{-1}(D_{H}-\gamma D)+H^{-2}(D-1)^2(\alpha D+\beta D^{-1})+\delta \psi   \\
\hline \\
D_{VE} - model &  D_{\rho\rho}=\phi D_{\rho}^2-\rho^{-1}(D_{\rho}-\gamma D)+\rho^{-2}(D-1)^2(\alpha D+\beta D^{-1})+\delta \psi  \\
\hline \\
D_{VF} - model &  D_{pp}=\phi D_{p}^2-p^{-1}(D_{p}-\gamma D)+p^{-2}(D-1)^2(\alpha D+\beta D^{-1})+\delta \psi  \\
\hline
\end{array} \end{equation}
Here $\phi=0.5D^{-1}+(D-1)^{-1}, \quad \psi= D(D+1)(D-1)^{-1}$.
\vspace{1cm}\\
 6) {\bf $D_{VI}$ - models.}
  \\
  \begin{equation}
 \begin{array}{|c|c|}
\hline \\
D_{VIA} - model &  \ddot{D}=\varphi(t)\dot{D}^2-\xi(t)\dot{D}-\frac{D(D-1)(D-t)}{t^{2}(t-1)^{2}}[\alpha +\frac{\beta t}{ D^{2}}+\zeta(t)]  \\
\hline \\
D_{VIB} - model &  D_{aa}=\varphi(a)D_{a}^2-\xi(a) D_{a}-\frac{D(D-1)(D-a)}{a^{2}(a-1)^{2}}[\alpha +\frac{\beta a}{ D^{2}}+\zeta(a)]  \\
\hline \\
D_{VIC} - model &  D_{NN}=\varphi(N) D_{N}^2-\xi(N) D_{N}-\frac{D(D-1)(D-N)}{N^{2}(N-1)^{2}}[\alpha +\frac{\beta N}{ D^{2}}+\zeta(N)] \\
\hline \\
D_{VID} - model &  D_{HH}=\varphi(H) D_{H}^2-\xi(H) D_{H}-\frac{D(D-1)(D-H)}{H^{2}(H-1)^{2}}[\alpha +\frac{\beta H}{ D^{2}}+\zeta(H)] \\
\hline \\
D_{VIE} - model &  D_{\rho\rho}=\varphi(\rho) D_{\rho}^2-\xi(\rho) D_{\rho}-\frac{D(D-1)(D-\rho)}{\rho^{2}(\rho-1)^{2}}[\alpha +\frac{\beta \rho}{ D^{2}}+\zeta(\rho)] \\
\hline \\
D_{VIF} - model &  D_{pp}=\varphi(p) D_{p}^2-\xi(p) D_{p}-\frac{D(D-1)(D-p)}{p^{2}(p-1)^{2}}[\alpha +\frac{\beta p}{ D^{2}}+\zeta(p)]\\
\hline
\end{array} \end{equation}
Here
\begin{eqnarray}
		\varphi(z)&=&0.5[D^{-1}+(D-1)^{-1}+(D-z)^{-1}],\\
	\xi(z)&=& z^{-1}+(z-1)^{-1}+(D-z)^{-1},\\
 \zeta(z)&=&\gamma(z-1)(D-1)^{-2}+\delta z(z-1)(D-z)^{-2}.\end{eqnarray}
\subsection{Nonintegrable D - models}
Nonintegrable D - models can be induced by some ODE's which are nonintegrable e.g. as in our previous papers \cite{Kuralay1}-\cite{Gulia}.  
\subsection{Solutions}
It is important that all above presented  D - models admit exact solutions.  Let's consider here one   example.
Let $D(p, \rho)$ has the form
\begin{equation}
D=0.5(p+\rho).
 \end{equation}
Consider the D$_{VID}$ - model [see the system  (5.6)].  This model in fact has the form
	\begin{eqnarray}
		p&=&-2\dot{H}-3H^2,\\
	\rho&=&3H^2,\\
 D_{HH}&=&\varphi(H) D_{H}^2-\xi(H) D_{H}-\frac{D(D-1)(D-H)}{H^{2}(H-1)^{2}}[\alpha +\frac{\beta H}{ D^{2}}+\zeta(H)]\\\dot{\rho}&=&-3H(\rho+p),
	\end{eqnarray}
Our aim is solve this system. For the particular case (5.10), the equation (5.13) has the following particular solution \cite{Ablowitz}
 \begin{equation}
D=\kappa H^2, \quad (\kappa=const).
 \end{equation}
 From (5.11)-(5.12) we get
  \begin{equation}
\dot{H}=-\kappa H^2.
 \end{equation}
 which has the solution
  \begin{equation}
H=\frac{1}{\kappa(t-t_0)}.
 \end{equation}
  	\section{Artificial two-dimensional models}
 	FRW cosmological models that we considered above are one-dimensional. But for some reason we would like to have two-dimensional cosmological models in FRW spacetime. And there are (may be) no legal ways to construct two-dimensional models starting from one-dimensional models. If so  let's try to use the "illegal" ways e.g.  introducing the artificial  "coordinate". As such artificial  coordinate we can use e.g.  one of  physical parameters of the original model e.g. the cosmological constant ($\Lambda$). So we have may be two coordinates: one legal coordinate - $t$ (time) and one "illegal" coordinate - $\Lambda$. Now we are ready to write our artificial two-dimensional cosmological models in FRW spacetime.  
 	
 	1) As an example, consider the following model e.g. for the scale factor [$a_{\Lambda}=da/d\Lambda, \quad a_t=da/dt$ etc]:
 	\begin{equation}
a_t=0.75(a^2)_{\Lambda}
 \end{equation}
 or its twin
 	\begin{equation}
a_{\Lambda}=0.75(a^2)_{t}
 \end{equation}
 which are known to develop shocks [{\bf Exercise 1}: \textit{What means these shocks for the dynamics of the universe?}]. Eqs. (6.1) and (6.2)  are nothing but the dispersionless Korteweg-de Vries equations (dKdVE)  	 or Riemann equations. It is well-known that the equations (6.1) and (6.2) have the following solutions 
 \begin{equation}
a(\Lambda, t)=h(\Lambda-0.75a t)
 \end{equation}
 and\begin{equation}
a(t, \Lambda)=h(t- 0.75a\Lambda),
 \end{equation}	 	
 respectively. Here $h$ is an arbitrary function. Also we see e.g. from the solution (6.3) that the velocity of a point of the wave, with constant amplitude $a$, is proportional to its amplitude leading to the "breaking" of the wave. Also we note that the wave also develops  discontinuities in its evolution [{\bf Exercise 2}: \textit{What means these discontinuities for the dynamics (evolution) of the universe?}]. Let's now we give the following particular solutions of the equations (6.1) and (6.2):
  \begin{equation}
a=(\beta_1 +\beta_2\Lambda)(\beta_3-1.5\beta_2 t)^{-1}
 \end{equation}
 and\begin{equation}
a=(\beta_1 +\beta_2 t)(\beta_3-1.5\beta_2 \Lambda)^{-1},
 \end{equation}	 	
 respectively, where $\beta_i=consts$.
 
 2) Our second example is the Euler-Tricomi equation
 \begin{equation}
a_{tt}=ta_{\Lambda\Lambda}.
 \end{equation}
 It is known that this equation has the following particular solution
  \begin{equation}
a=\alpha(3\Lambda^2+t^3)+\beta(\Lambda^3+t^3\Lambda)+\delta(6t\Lambda^2+t^4).
 \end{equation}
 The twin of the equation (6.7) reads as 
 	\begin{equation}
a_{\Lambda\Lambda}=\Lambda a_{tt}.
 \end{equation}
 
 3) Let's now present  the Dym equation
 \begin{equation}
a_t=(a^{-0.5})_{\Lambda\Lambda\Lambda}
 \end{equation}
 and its twin
 	\begin{equation}
a_{\Lambda}=(a^{-0.5})_{ttt}.
 \end{equation}
 
 4) Our  next example is given by \cite{Strachan}
 \begin{equation}
(\ln a)_{tt}=a_{\Lambda\Lambda}
 \end{equation}
 and its twin
 	\begin{equation}
(\ln  a)_{\Lambda\Lambda}=a_{tt}.
 \end{equation}
 
 5) Our last example  is given by 
 \begin{equation}
a_{t}=\mu a_{\Lambda\Lambda}.
 \end{equation}
It is the heat equation  and  it is well-known that it  has the following fundamental solution [as $a(\Lambda, t=0)=\delta(\Lambda)$]
\begin{equation}
a=(4\pi \mu t)^{-0.5}e^{-0.25\mu^{-1}t^{-1}\Lambda^2}.
 \end{equation}
Note that the  twin of the equation (6.14) looks like
 	\begin{equation}
 a_{\Lambda}=\mu a_{tt}.
 \end{equation}
 
 \section{Conclusion}

It is important  to study the physical and mathematical nature of FRW cosmological models as they play a crucial role in modern cosmology. In this work,  a new classes integrable and nonintegrable FRW cosmological models   were proposed. To construct integrable models we implant integrable equations, in our Painlev$\acute{e}$ equations, into the body of the original gravitational equations. It is remarkable that all proposed integrable models admit exact solutions. For some of them, exact solutions are presented. Finally in Appendix, we extend  the obtained results for g-essence models and its two reductions: k-essence and f-essence. 

\section{Appendix. Integrable and nonintegrable g-essence models and their  k-essence and f-essence reductions} For g-essence we have
 	\begin{equation}
 p=K, \quad \rho=2XK_X+YK_Y-K,
 \end{equation}
where $K$ is the Lagrangian for g-essence, $X$ and $Y$ are the kinetic terms for the scalar and spinor fields, respectively. Below we give some examples of integrable models for g-essence.
  \vspace{1cm}\\
 2) {\bf $E_{I}$ - models.}\\ 
 \begin{equation}
 \begin{array}{|c|c|}
\hline \\
E_{IA} - model &  \ddot{K}=6K^2+t   \\
\hline \\
E_{IB} - model &  K_{aa}=6K^2+a \\
\hline \\
E_{IC} - model &  K_{NN}=6K^2+N  \\
\hline \\
E_{ID} - model &  K_{HH}=6K^2+H \\
\hline \\
E_{IE} - model &  K_{XX}=6K^2+X \\\hline \\
E_{IF} - model &  K_{YY}=6K^2+Y \\
\hline
\end{array} \end{equation}
\vspace{1cm}\\
 2) {\bf $E_{II}$ - models.}\\ 
  \begin{equation} \begin{array}{|c| c|}
\hline \\
E_{IIA} - model &  \ddot{K}=2K^3+tK+\alpha   \\
\hline \\
E_{IIB} - model &  K_{aa}=2K^3+aK+\alpha  \\
\hline \\
E_{IIC} - model &  K_{NN}=2K^3+NK+\alpha   \\
\hline \\
E_{IID} - model &  K_{HH}=2K^3+HK+\alpha  \\
\hline \\
E_{IIE} - model &  K_{XX}=2K^3+XK+\alpha  \\
\hline \\
E_{IIF} - model &  K_{YY}=2K^3+YK+\alpha  \\
\hline
\end{array} \end{equation}
 3) {\bf $E_{III}$ - models.} \\
  \begin{equation}
 \begin{array}{|c|c|}
\hline \\
E_{IIIA} - model &  \ddot{K}=K^{-1}\dot{K}^2-t^{-1}(\dot{K}-\alpha K^2-\beta)+\gamma K^3+\delta K^{-1}  \\
\hline \\
E_{IIIB} - model &  K_{aa}=K^{-1}K_{a}^2-a^{-1}(K_{a}-\alpha K^2-\beta)+\gamma K^3+\delta K^{-1} \\
\hline \\
E_{IIIC} - model &  K_{NN}=K^{-1}K_{N}^2-N^{-1}(K_{N}-\alpha K^2-\beta)+\gamma K^3+\delta K^{-1}  \\
\hline \\
E_{IIID} - model &  K_{HH}=K^{-1}K_{H}^2-H^{-1}(K_{H}-\alpha K^2-\beta)+\gamma K^3+\delta K^{-1} \\
\hline \\
E_{IIIE} - model &  K_{XX}=K^{-1}K_{X}^2-X^{-1}(K_{X}-\alpha K^2-\beta)+\gamma K^3+\delta K^{-1} \\
\hline \\
E_{IIIF} - model &  K_{YY}=K^{-1}K_{Y}^2-Y^{-1}(K_{Y}-\alpha K^2-\beta)+\gamma K^3+\delta K^{-1} \\
\hline
\end{array} \end{equation}
\vspace{1cm}\\
 4) {\bf $E_{IV}$ - models.} 
  \\
  \begin{equation}
 \begin{array}{|c|c|}
\hline \\
E_{IVA} - model &  \ddot{K}=0.5K^{-1}\dot{K}^2+1.5K^3+4tK^2+2(t^2-\alpha)K+\beta K^{-1}  \\
\hline \\
E_{IVB} - model &  K_{aa}=0.5K^{-1}K_{a}^2+1.5K^3+4aK^2+2(a^2-\alpha)K+\beta K^{-1}\\
\hline \\
E_{IVC} - model &  K_{NN}=0.5K^{-1}K_{N}^2+1.5K^3+4NK^2+2(N^2-\alpha)K+\beta K^{-1} \\
\hline \\
E_{IVD} - model &  K_{HH}=0.5K^{-1}K_{H}^2+1.5K^3+4HK^2+2(H^2-\alpha)K+\beta K^{-1}\\

\hline \\
E_{IVE} - model &  K_{XX}=0.5K^{-1}K_{X}^2+1.5K^3+4XK^2+2(X^2-\alpha)K+\beta K^{-1}\\
\hline \\
E_{IVD} - model &  K_{YY}=0.5K^{-1}K_{Y}^2+1.5K^3+4YK^2+2(Y^2-\alpha)K+\beta K^{-1}\\
\hline
\end{array} \end{equation}
\vspace{1cm}\\
 5) {\bf $E_{V}$ - models.} 
  \\
  \begin{equation}
 \begin{array}{|c|c|}
\hline \\
E_{VA} - model &  \ddot{K}=\phi\dot{K}^2-t^{-1}(\dot{K}-\gamma K)+t^{-2}(K-1)^2(\alpha K+\beta K^{-1})+\delta \psi  \\
\hline \\
E_{VB} - model &  K_{aa}=\phi K_{a}^2-a^{-1}(K_{a}-\gamma K)+a^{-2}(K-1)^2(\alpha K+\beta K^{-1})+\delta \psi  \\
\hline \\
E_{VC} - model &  K_{NN}=\phi K_{N}^2-N^{-1}(K_{N}-\gamma K)+N^{-2}(K-1)^2(\alpha K+\beta K^{-1})+\delta \psi   \\
\hline \\
E_{VD} - model &  K_{HH}=\phi K_{H}^2-H^{-1}(K_{H}-\gamma K)+H^{-2}(K-1)^2(\alpha K+\beta K^{-1})+\delta \psi   \\
\hline \\
E_{VE} - model &  K_{XX}=\phi K_{X}^2-X^{-1}(K_{X}-\gamma K)+X^{-2}(K-1)^2(\alpha K+\beta K^{-1})+\delta \psi   \\
\hline \\
E_{VD} - model &  K_{YY}=\phi K_{Y}^2-Y^{-1}(K_{Y}-\gamma K)+Y^{-2}(K-1)^2(\alpha K+\beta K^{-1})+\delta \psi   \\
\hline
\end{array} \end{equation}
Here $\phi=0.5D^{-1}+(K-1)^{-1}, \quad \psi= K(K+1)(K-1)^{-1}$.
\vspace{1cm}\\
 6) {\bf $E_{VI}$ - models.}
  \\
\\
  \begin{equation}
 \begin{array}{|c|c|}
\hline \\
E_{VIA} - model &  \ddot{K}=\varphi(t)\dot{K}^2-\xi(t)\dot{K}-\frac{K(K-1)(K-t)}{t^{2}(t-1)^{2}}[\alpha +\frac{\beta t}{ K^{2}}+\zeta(t)]  \\
\hline \\
E_{VIB} - model &  K_{aa}=\varphi(a)K_{a}^2-\xi(a) K_{a}-\frac{K(K-1)(K-a)}{a^{2}(a-1)^{2}}[\alpha +\frac{\beta a}{ K^{2}}+\zeta(a)]  \\
\hline \\
E_{VIC} - model &  K_{NN}=\varphi(N) K_{N}^2-\xi(N) K_{N}-\frac{K(K-1)(K-N)}{N^{2}(N-1)^{2}}[\alpha +\frac{\beta N}{ K^{2}}+\zeta(N)] \\
\hline \\
E_{VID} - model &  K_{HH}=\varphi(H) K_{H}^2-\xi(H) K_{H}-\frac{K(K-1)(K-H)}{H^{2}(H-1)^{2}}[\alpha +\frac{\beta H}{ K^{2}}+\zeta(H)] \\
\hline \\
E_{VIE} - model &  K_{XX}=\varphi(X) K_{X}^2-\xi(X) K_{X}-\frac{K(K-1)(K-X)}{X^{2}(X-1)^{2}}[\alpha +\frac{\beta X}{ K^{2}}+\zeta(X)] \\
\hline \\
E_{VIF} - model &  K_{YY}=\varphi(Y) K_{Y}^2-\xi(Y) K_{Y}-\frac{K(K-1)(K-Y)}{Y^{2}(Y-1)^{2}}[\alpha +\frac{\beta Y}{ K^{2}}+\zeta(Y)] \\
\hline
\end{array} \end{equation}

 \end{document}